\documentclass[prl,twocolumn,showpacs,amsmath,amssymb]{revtex4}
\usepackage{amssymb}
\usepackage[dvips]{graphicx}

\begin{document}

\title{Phase slip phenomena in superconductors: from ordered to chaotic dynamics}
\author{ Mathieu Lu-Dac and V. V. Kabanov}

\affiliation{Jozef Stefan Institute, Jamova 39, 1001 Ljubljana,
Slovenia}

\begin{abstract}
We consider flux penetration to a 2D superconducting cylinder.
We show that in the low field limit the kinetics is deterministic.
In the strong field limit the dynamics becomes stochastic.
Surprisingly the inhomogeneity in the cylinder reduces the level
of stochasticity because of the predominance of Kelvin-Helmholtz
vortices.
\end{abstract}

\maketitle


The kinetics of vortex production in superconductors and
superfluids is  one of the intriguing problems of condensed
matter physics. It is interesting not only in the field
of solid state physics but represents as well a very good
model to study topological phase transitions in cosmology and
other branches of physics \cite{zurek1}. In
the last few decades different scenarios for vortex production
in superconductors and superfluids were proposed. The most common
way to produce vortices is to increase the superfluid velocity in
order to reduce the energy barrier between homogeneous flow and
flow with vortices. This mechanism is observed in rotating
$^{3}\text{He}$ where vortex nucleation and critical velocities
are measured \cite{parts}. In 2D homogeneous superconducting films,
increasing the current leads to dynamics which are similar
to the phase slip (PS) transition in 1D \cite{ludac}.  The order
parameter (OP) reaches zero along a straight line across the film
and the phase displays a $2\pi$ jump along this line. This PS
line solution \cite{Weber09} corresponds to the
deterministic and most ordered PS kinetics in 2D. The
inhomogeneity caused by current contacts leads to a qualitatively
similar picture. The OP is strongly suppressed along a straight
line across the film but it reaches zero only at two points on
this line. This pair of vortices is called kinematic vortex-antivortex
(VaV) pair \cite{andronov93}. It spreads quickly in
opposite directions along this line propagating the $2\pi$ jump of
the phase. Therefore PS occurs without formation of well defined
VaV pairs \cite{berdiyorov}.

A different scenario of vortex production was proposed by
Kibble\cite{kibble1} and Zurek\cite{zurek2} (KZ). When the sample
is quickly quenched through the critical temperature $T_c$, the
nucleation of the low temperature phase starts in different places
with uncorrelated phases of OP. Then, domains grow and start to
overlap leading to the formation of vortices. This mechanism is a
promising way to test cosmological theories in condensed matter
physics \cite{Ruutu96-98, Maniv03}. This dynamics is stochastic and
sensitive to small variations of initial conditions. On the
other hand the dependence of the vortex density on the quench time
and their spatial correlation are universal. Later, in Refs.
\cite{kibblevolovik,kopninthuneberg}, it was proposed that the
quench occurs not only due to fast temperature change but also due
to the temperature front propagation. Aranson \textit{et al.}
considered the case of a temperature quench in the presence of
external current \cite{aranson}. The new phase with zero current
grows after the quench. Therefore on the border of the quenched
region, the superfluid velocity has tangential discontinuity,
leading to vortex formation, similarly to the classical
hydrodynamic Kelvin-Helmholtz (KH) instability \cite{fridman}
which is also known in superfluids \cite{Blaauwgeers02,volovikKH}.
Moreover, the KH instability suppresses the development of KZ
vortices \cite{aranson}. In this paper, kinematic VaV, KZ and KH
vortices are distinguished by their production mechanism although they
are topologically equivalent.

To demonstrate how deterministic type of dynamics becomes
stochastic, we model a superconducting film rolled on a cylinder
in an external time dependent magnetic field parallel to the
cylinder axis (Fig.1). Depending on the applied magnetic
field and the dimensions of the ring, we follow the evolution
from the deterministic PS line dynamics to the stochastic behavior
described by the KZ mechanism. In the proposed model, topological
defects are generated by the intrinsic quench induced by the
external field. The evolution towards
stochastic behavior is strongly influenced by the KH instability
which develops in the presence of inhomogeneities.  To model the
inhomogeneity of the film we assume that there is a thin stripe
of superconductor along the film with a different coherence length.
\begin{figure}
\includegraphics[width = 40mm]{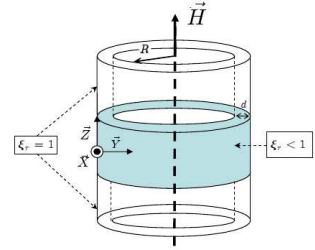}
\caption{Geometry of the system :  a 2D cylinder with an applied
magnetic field $\mathbf{H}$.}
\end{figure}
The thickness of the film $d$ is small $d\ll \xi\
\ll \lambda_{\text{eff}}$. Here $\xi$ is the coherence length and
$\lambda_{\text{eff}}$ is the Pearl penetration depth. Therefore
we can neglect all corrections to the external magnetic field
$\mathbf{H}$ caused by the current in the film. The radius of the
film is $R > \xi$.
The time dependent Ginzburg-Landau (TDGL) equation in
dimensionless units has the form:
\begin{equation}
u(\frac{\partial\psi}{\partial t}+i\Phi \psi) =
b(z)(\psi-\psi|\psi|^{2})-(i\nabla+\mathbf{a})^{2}\psi
+\eta.\label{tdgl1}
\end{equation}
Here $\psi$ is the dimensionless complex OP, the spacial
coordinate ${\mathbf r}$ is measured in units of $\xi$ and  time
is measured in units of phase relaxation time
$\tau_{\theta}=\frac{4\pi\lambda_{\text{eff}}\sigma_{n} }
{c^{2}}$, $\lambda_{\text{eff}}=\frac{\lambda^{2}}{d}$, $\lambda$
is the bulk penetration depth, $\sigma_{n}$ is the normal state
conductivity, and $c$ is the speed of light. The parameter
$u=\frac{\tau_{\psi}}{\tau_{\theta}}$ is a material dependent
parameter, where $\tau_{\psi}$ is the relaxation time of the
amplitude of the OP. According to the microscopic theory, $u$ is
ranging from $5$ to $12$ but we assume $0<u<\infty$. The vector
potential $\mathbf{a}$ is measured in units of
$\frac{\phi_{0}}{2\pi\xi}$ where $\phi_{0}$ is the flux quantum.
The function $b(z)$ models the $z$ dependence of the coherence
length $\xi(z)$. As shown in Fig.1, we chose
$\xi(z)=\xi/\sqrt{b(z)}$ and
$b(z)=1-b^{2}\vartheta(z-w/4)\vartheta(3w/4-z)$. Here $w$ is the
width of the film in units of $\xi$, $b$ parameterizes the level
of inhomogeneity of the film and $\vartheta(x)$ is the Heaviside
step function. Here we use periodic boundary conditions and the
boundary condition with vacuum \cite{degennes} at $z=0$
and $z=w$. The equation for the electrostatic potential $\Phi$,
measured in units of $\frac{\phi_{0}}{2\pi c\tau_{\theta}}$,where
$e$ is the electronic charge and $\hbar$ is the Planck constant,
reads:
\begin{equation}
\nabla^{2}\Phi=-\nabla\Bigl[
{{i}\over{2}}(\psi^{*}\nabla\psi-\psi\nabla\psi^{*})
+\mathbf{a}|\psi|^{2}\Bigr].\label{tdgl2}
\end{equation}
To model the process of vortex formation we assume that at time
$t<0$ external magnetic field is absent.
At $t=0$ the field suddenly appears and stays constant for $t>0$
i.e. tangential component of the vector potential is
$a\vartheta(t)$. We thus study the kinetics of the vortex
generation as a function of $a$ with different values of $u$.

Let us first consider the stability of the solution in the uniform
case. We linearize the TDGL Eqs.(\ref{tdgl1},\ref{tdgl2}) in small
fluctuations of OP $f(\mathbf{r},t)=\psi(\mathbf{r},t)-\psi_{0}$
and search for a solution in the form
$f(\mathbf{r},t)=\sum_{\mathbf{k}}C_{\mathbf{k}}
\exp{(i\mathbf{kr}+\lambda_{\mathbf{k}} t)}$. It is clear that the
transverse $k_{z}$ component always contributes to the stability
of the initial state. Therefore, the condition
$\lambda_{\mathbf{k}}>0$ is the same as in 1D \cite{ludac}:
$\frac{\phi}{\phi_0}
> > \frac{R}{\xi\sqrt{3}}$, where $\phi$ is the magnetic flux
through the ring at $t>0$. This condition provides a rough
estimate for the number of the expected PS events $N\sim
\frac{\phi}{\phi_0}$. It defines the first critical value of the
external field $a_{c1}=1/\sqrt{3}$. Therefore, in the low field
limit $a_{c1}\leq a \leq 1$, the dynamics will be similar to the 1D
case with very weak $z$-dependence. Well defined vortices may
appear in this region of the field if the film is inhomogeneous as
shown in Fig.1. The situation is different when the field
$a$ increases further. Dropping $k_{z}=0$, the
eigenvalues are:
\begin{eqnarray}
\lambda^{(1,2)}_{\mathbf{k}}=-\psi_{0}^{2}/2+(1
-2\psi_0^2-a^2-k^2)/u \nonumber \\ \pm \sqrt{(16\psi_0^2
a^2+\psi_0^4(u-2)^2+16k^2a^2)/4u^2} \label{decay}
\end{eqnarray}
$\lambda_{\mathbf{k}=0}=\frac{1-2
\psi_0^2-a^2}{u}-\frac{\psi_{0}^2}{2} - \sqrt{(16\psi_0^2
a^2+\psi_0^4(u-2)^2)/4u2}$ describes the decay rate of the
uniform solution. On the other hand for finite $k$,
$\lambda_{\mathbf{k}}$ is positive and characterizes the growth of
the corresponding Fourier components $C_{\mathbf{k}}$. The fastest
growth is found for $k=\frac{1}{4a} \sqrt{-16u\psi_0^2a^2-
\psi_0^4(u-2)^2+16a^4}$ and the rate is
determined by
$\lambda_{\text{max}}=\frac{1}{16ua^2}(8(u-4)\psi_0^2a^2
+16a^2+\psi_0^4(u-2)^2)$. The qualitative
difference in kinetics takes place when the decay rate of the
uniform solution becomes faster then the growth of the new phase.
This effect is similar to the quench through $T_{c}$ in the KZ
mechanism \cite{zurek1,kibble1}. We find that at
$a>a_{c2}=\sqrt{2}$, the OP is suppressed to zero and the growth
of the phase with finite $k$ is accompanied by the rapid
development of vortices. The density of vortices may be estimated
using Zurek arguments where the quench time should be replaced by
$\tau_{Q}=(a^{2}-1)^{-1}$ leading to $n\propto \tau_{Q}^{-1/2}$
\cite{zurek3}.

We simulate Eqs.(\ref{tdgl1},\ref{tdgl2}) using the fourth order
Runge-Kutta method. The spatial derivatives are evaluated using
a finite difference scheme of
second order or using a fast fourier transform algorithm depending
on the boundary conditions. The choice of the algorithm is made to
optimize the convergence and the calculation times. The
calculations are performed for the vector potential $0<a<5$ and
for the total flux $\phi$ through the ring ranging from $0$ to $50
\phi_0$.

\begin{figure*}
           {\includegraphics[width=175mm]{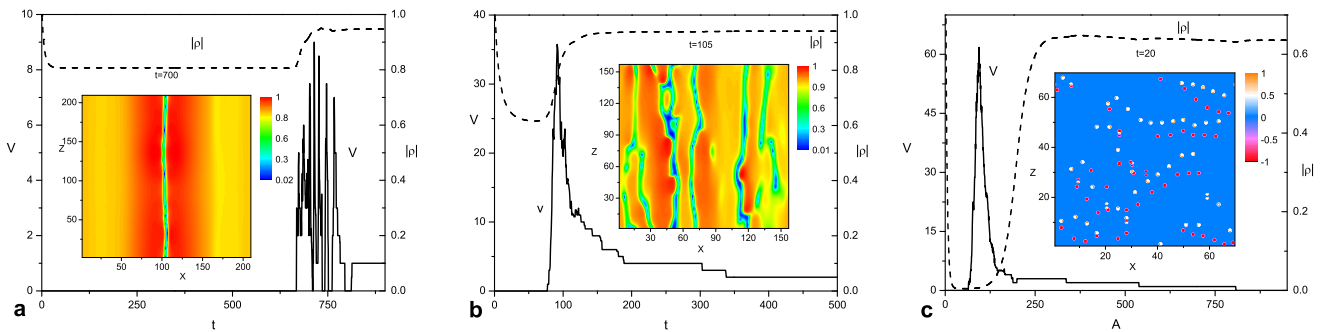}}
           \caption{(Color online) Total number of vortices $v$ in the system
           and the sample average
           value $|\rho|$ of the OP as a function of time for
           $\phi/\phi_0=20$ and $a=0.6$ (a), $a=0.8$(b) and $a=1.8$ (c).
           The insets represent snapshots the amplitude of OP at
           $t=700$ (a) and $t=105$ (b). For the quenched case (c),
           the snapshot displays the local vorticity at $t=20$. }
\end{figure*}

We investigate the flux penetration into the homogeneous ring for
two different boundary conditions. In the case of periodic
boundary conditions we identify different regimes in accordance
with Eq.(\ref{decay}). In the small field limit $a<a_{c1}$ the
ring is in a stable state and the penetration of the magnetic flux
into the ring can only be induced by a very strong noise $\eta$ in
Eq.(\ref{tdgl1}). When $a_{c1}<a<a_{c2}$, in agreement with the
stability analysis, the PS kinetics depends on the external
magnetic field. When $\frac{\phi}{\phi_0}<10$ the kinetics is
similar to the 1D case. The transition is characterized by one or
more lines in the $z$-direction where the OP decreases to zero
(the PS line case \cite{andronov93}). These lines may appear
simultaneously or consecutively in time, depending on $u$
\cite{ludac}. As expected, the number of PS events is determined
by the ratio $\frac{\phi}{\phi_0}$. These PS lines represent the
limiting case of kinematic VaV pairs travelling with infinite
velocity.

When the flux is increased ($\frac{\phi}{\phi_0}>10$), the
kinematic vortices become clearly distinguishable. In Fig.2(a), we
present the time evolution of the average value of the OP together
with the time dependence of the number of vortices in the
sample. The kinetics is characterized by series of consecutive PS
events well separated in time (Fig.2(a)). As it was noticed
\cite{vodo07}, few VaV pairs may propagate along the same line at
the same time.  PS  events are produced by kinematic VaV pairs
propagating along the same line where the amplitude of OP is
reduced. Kinematic vortices can propagate in the same direction,
one after another or in opposite direction leading to annihilation
of VaV pairs and accelerating the dynamics. Contrary to
\cite{berdiyorov}, kinematic VaV pairs are formed without any
inhomogeneity in the film. At higher fluxes kinematic VaV pairs
are randomly created on the line like in the case of a "1D
quench". In the $x$ direction, the dynamics remains very ordered
with values of the standard deviation of the position of the
vortices $\sqrt{\bar{\delta x^{2}}}$ approaching $0.5\xi$.

\begin{figure*}
           {\includegraphics[width=175mm]{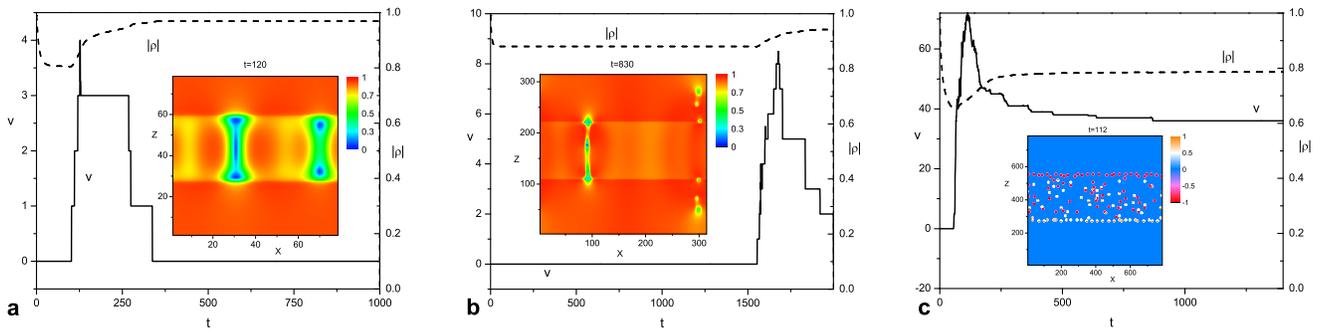}}
           \caption{(Color online) Total number of vortices $v$ in the system and
           the sample average value $|\rho|$ of the OP as a function
           of time for $a1=0.4$, $\phi/\phi_0=5$ and $b=4$ (a); $\phi/\phi_0=20$
           and $b=2.25$ (b); $\phi/\phi_0=50$ and $b=12.25$ (c). The insets
           represent snapshots of the amplitude of the OP at $t=120$ (a) ;
           $t=790$ and $t=830$ (b). For the quasi quenched case (c), the snapshot
           displays the local vorticity at $t=112$. }
\end{figure*}

With the further increase of $a$ the number of PS lines increases
and the kinetics becomes more stochastic because of the
interaction of different PS lines. As a result, straight lines are
replaced by vortex rivers which become broader and have finite
curvature (Fig.2(b)). The vortex rivers are comparable to the vortex self-organization
discussed in \cite{Aranson96} under different boundary conditions.
Along one vortex river, few vortex-antivortex pairs are
propagated. The kinetics is determined by the motion of these
pairs along the rivers and finally by their annihilation.
Importantly, the sample average of OP never reaches zero,
contrarily to the case of the large field $a>a_{c2}$. The total
number of vortices in the beginning of the process is larger than
$\frac{\phi}{\phi_0}$ (Fig.2(b)) which is also an indication of
the growing importance of chaotic behavior in the dynamics. The
values of $\sqrt{\bar{\delta x^{2}}}$ are also strongly enhanced,
reaching $2\pi R/3$. The velocity of vortices along the rivers
becomes smaller which is seen from the time dependence of the
vortex number (Fig.2(d)). Nevertheless the velocity is still high
compared to the case when the OP has recovered to its equilibrium
value. The last regime $a>a_{c2}$ is presented in Fig.2(c). Here
the quench condition is satisfied and the OP decreases uniformly
until it reaches zero (Fig. 2(c)). As a result, the new phase
starts to grow uncorrelated and the vortices are created
randomly. The number of vortices is substantially larger than
$\frac{\phi}{\phi_0}$. Most of these vortices recombine rapidly.
The remaining vortices move slowly through the sample propagating
the $2\pi$ phase jump. The random dispersion of these vortices is
a fingerprint of the KZ mechanism. Indeed, $\sqrt{\bar{\delta
x^{2}}}$ reaches now $2\pi R/2$, which means that vortex
distribution is completely random. Another characteristic of the
KZ scenario is that the vortices are created while the order
parameter is very close to zero and not during the fast growth
like in the previous cases as one can see by comparing Fig. 2(a)
and (b) with Fig. 2(c). It is important to notice that the total
net vorticity is strictly equal to zero at any time in the case of
periodic boundary condition in the z direction.

For vacuum boundary conditions \cite{degennes} the
kinetics is very similar. When $a_{c1}<a<a_{c2}$ and
$\frac{\phi}{\phi_0}<10$ one or more lines with reduced OP are
formed. The difference is that the PS lines here have finite
curvature, because they start to grow from the edges of the film
and finally connect each other. Further increase of the flux,
keeping $a$ constant leads to the formation of flux rivers. The
most important difference is that not all "rivers" necessarily
connect two edges of the film. As a result some of them ended in
the middle of the film, leading to the relatively small vorticity.
These remaining vortices and antivortices propagate slowly to the
edges of the film and kinetic is determined by the slow vortex
motion. The dynamics when $a>a_{c2}$ is governed by KZ mechanism,
as in the previous case but the total net vorticity may be finite.

In the case of an inhomogeneous superconductor, the effective
coherence length is now $z$-dependent $\xi(z) = \xi/\sqrt{b(z)}$.
Therefore only the middle part of the ring may be unstable while
the other parts of the film remain in the metastable state. The
introduction of $z$-dependence of the parameters in
Eq.(\ref{tdgl1}) is designed to enhance the transverse vortex
dynamics and allows to demonstrate different mechanisms of vortex
formation. As expected, the PS dynamics starts first in the region
with stronger current and is characterized by $a$ and
$\frac{\phi}{\phi_0}$.

In the region $a_{c1}<a<a_{c2}$ and small flux
$\frac{\phi}{\phi_0}<10$ the initial stage of the kinetics is
similar to kinetics in the homogeneous film. The VaV pairs are not
well defined. However, when kinematic VaV pairs approach the low
current regions, they become well defined and are slowing down
(fig.3(a)). Therefore vortices are stabilized near the line where
the tangential velocity has discontinuity. These vortices
represent another case of KH instability in superconductors. This
instability leads to the formation of well defined vortices and
governs the kinetics of the PS. To the best of our knowledge this
is the only instability which allows vortex production in the low
flux limit.

When the flux through the ring is large $\frac{\phi}{\phi_0}>10$,
the initial fast dynamics is similar to the dynamics in the
homogeneous case until vortices reach the low current regions.
Then they become slow and well defined. As it is seen in
Fig.3(b), the vortices propagate one after another to the film
edge, demonstrating the vortex-vortex attraction even in the case
when the OP has already recovered.

The further increase of $a>a_{c2}$ leads to the quench in the
middle part of the film (Fig.1). During the quench many KZ vortices
are created. Most of them are annihilated on a very short
time scale. The rest reaches the line separating the region with
different currents. The vortices almost stop near this line. The
further dynamics is determined by the diffusion of these vortices
to the film edges. When $a$ is large enough, the KH vortices
become well defined before the recovery of the OP in the middle
part of the film and therefore the inhomogeneity suppresses the KZ
mechanism in agreement with Ref.\cite{aranson}, making kinetics
less stochastic.


Experimentally, observing such dynamics of  vortices might
be a real challenge because the short characteristic times does
not allow the use of instruments with sufficient space resolution.
However, recent works \cite{Maniv03, Silhanek10} showed that
freezing the dynamics can characterize both KZ and vortex river
scenarios. Another idea is to use time resolved femtosecond
optical spectroscopy as proposed in the Ref.\cite{yusupov}.
As it is shown in the Refs.\cite{ludac,suppl} the role of
heating is not important for the proposed
geometry of the film.

We have considered the kinetics of the flux penetration to the 2D
ring. We found out that for small values of the external field $a$,
the kinetics is
deterministic and essentially 1D. Increasing the flux $\phi$ creates
kinematic vortices and even leads to a 1D quench along the PS line
which is a first step towards stochastic behavior (see Ref.\cite{suppl}). Further
increase of $a$ leads to the formation of vortex rivers, and
ultimately to the quench of the sample leading to the stochastic dynamics of KZ vortices.
The dynamics in the inhomogeneous film demonstrates that the VaV pairs are the
topological analog of the PS mechanism in 2D but this analogy is
not as straightforward as is often believed. Finally,
our calculations for a partially quenched film indicate that KH
vortices at the interface are strongly predominant.

\end{document}


\title{Supplementary material for "Phase slip phenomena in superconductors: from ordered to chaotic dynamics"}
\author{ Mathieu Lu-Dac and V. V. Kabanov}
\affiliation{Jozef Stefan Institute, Jamova 39, 1001 Ljubljana,
Slovenia}

\maketitle

\section{Phase diagram of the different dynamics}

We observe different dynamics in the homogeneous case, depending on the value of the magnetic vector potential $a$ and the number of flux quanta $\phi/\phi_0$ penetrating the cylinder. In order to plot a phase diagram and display the different types of dynamics, we needed to choose between multiple criteria. We found out that the stochasticity is a good approach to this problem and chose the standard deviation of the $x$ coordinate of vortices to define regions where the dynamics are comparable. The normalized standard deviation $\frac{\sqrt{\bar{\delta x^{2}}}}{\pi R}$ is shown in Fig.1 (R is the radius of the cylinder). The different regions where the standard deviation is similar are separated by lines of constant standard derivation.
\begin{itemize}
\item Part $1$ of the phase diagram corresponds to the phase slip (PS) line solution. No vortices are present and we define there $\frac{\sqrt{\bar{\delta x^{2}}}}{\pi R}=0$ because the dynamics is very ordered.

\item Part $2$ ($0<\frac{\sqrt{\bar{\delta x^{2}}}}{\pi R}\leq 0.23$) corresponds to kinematic vortices traveling on a single line.

\item Part $3$ ($0.23<\frac{\sqrt{\bar{\delta x^{2}}}}{\pi R}<0.48$)  is assigned to multiple vortex rivers : stochasticity is increased by the the increasing number of rivers.

\item Part $4$ ($0.55<\frac{\sqrt{\bar{\delta x^{2}}}}{\pi R}$) corresponds to Kibble-Zurek (KZ) type dynamics which displays the highest possible stochasticity.

\item Part 5 ($0.48<\frac{\sqrt{\bar{\delta x^{2}}}}{\pi R}\leq 0.55$) represents the region where the standard deviation fails to clearly distinguish between KZ and vortex rivers. In this region, the area with low $a$ corresponds to vortex rivers and should belong to the part 3. Indeed, when the flux is increased at low $a$, the number of vortex river increases and the vortices often travel from one river to the other : their position becomes stochastic. On the opposite, at low flux, the area with high $a$ should belong to part 4.
\end{itemize}

\begin{figure}
\includegraphics[width = 70mm]{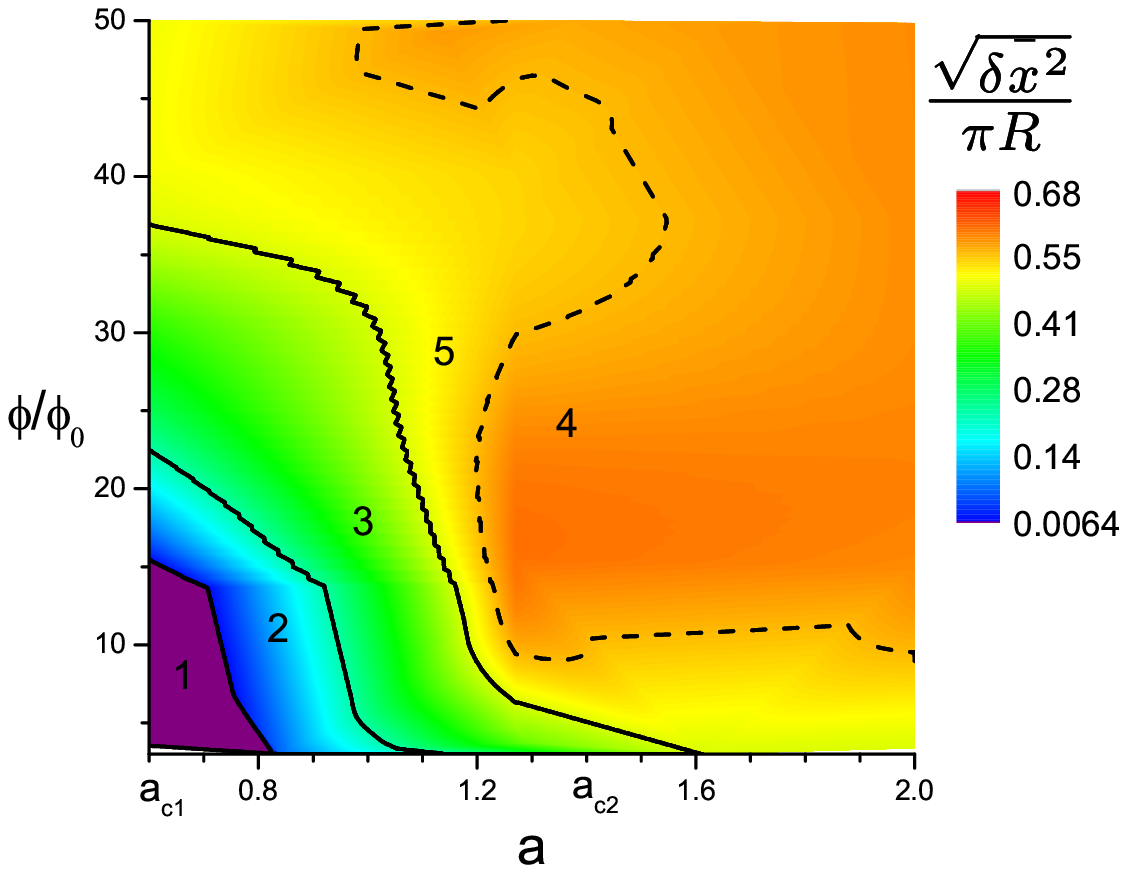}
\caption{(Color online) Phase diagram of the possible evolution drawn from the normalized standard deviation $\frac{\sqrt{\bar{\delta x^{2}}}}{\pi R}$ of the $x$ coordinate of the vortices. We chose to define $\frac{\sqrt{\bar{\delta x^{2}}}}{\pi R}=0$ when no vortices are present (Part 1). Part 1 corresponds to the PS line, part 2 the kinematic vortices, part 3 to vortex rivers and part 4 to KZ type dynamics. The part 5 represents a region where the standard deviation is not sufficient to distinguish between the dynamics. The standard deviation is plotted as a function of the vector potential $a$ and of the number of flux quanta (a) for $u=10$. $a_{c1}$ is the critical value under which dynamics are very improbable and $a_{c2}$ is the critical value over which we calculated KZ dynamics would occur. }
\end{figure}

\section{Heating}
The estimate of the heating of the cylinder may be done using Ohm's law. We consider that a part of the cylinder of area $\pi \xi^2$ for the vortex or $2\pi R \xi$ for the PS line, behaves like a normal conductor of conductivity $\sigma_n$ until its annihilation and sustains a current $J_s$ (in cgs units). The energy dissipated per unit of volume is, as in \cite{ludac}:
                \begin{equation}
                E=\frac{J_s^2}{\sigma_n}\tau_{\theta}=\frac{\left [\phi_0 a (1-a^2) \right ]^2}{16\pi^3 \xi^2\lambda_{\text{eff}}^2}.
                \end{equation}
                This energy is less than a quarter of the condensation energy $\frac{H_{c_2}^2}{16\pi\kappa^2}$. In the case of PS lines and kinematic vortices, when dynamics are fast and the areas behaving as normal conductors are small, the heat will be dissipated along the sample and through the contacts. It will not impact the dynamics. However, in the case of multiple vortex rivers or in the quenched KZ scenario, the total heating will be much larger but can be tackled by modern experimental techniques. Indeed, modern cooling methods are fast enough \cite{Maniv03}.